\documentclass[10pt, preprint,aps,prc,floatfix,
tightenlines,
nofootinbib,superscriptaddress]{revtex4-2}
\usepackage{graphicx,amsmath,amssymb,bm}
\usepackage{amsfonts}

\usepackage[utf8]{inputenc}
\usepackage{verbatim}
\usepackage{float}
\usepackage[labelfont={small},font=small,subrefformat=parens,caption=false]{subfig}
\usepackage{cancel}
\usepackage{multirow}
\usepackage{array}
\usepackage{xparse}
\usepackage{xspace}
\usepackage{xcolor}
\usepackage[inkscapeformat=pdf]{svg}

\usepackage{bigstrut}

\usepackage{physics}
\usepackage{color}
\usepackage{dsfont}
\usepackage{dcolumn}

\usepackage{xurl}
\usepackage[pdfencoding=auto, pdfpagelabels]{hyperref}

\usepackage[overload]{textcase}

\usepackage{bm}
\usepackage{enumitem}

\definecolor{linkcolor}{rgb}{0,0,0.40} 
\hypersetup{%
    pdfsubject=Paper,
    pdfkeywords= {Bayesian Model Mixing} {Bayesian methods} {software} {computational physics},
    unicode = true,
    breaklinks = true,
    colorlinks = true,
    linkcolor = linkcolor,
    citecolor = linkcolor,
    menucolor = linkcolor,
    urlcolor = linkcolor
}

\usepackage{cellspace}
\newcommand{\M}{\mathcal{M}}
\newcommand{\xvec}{\boldsymbol x}

\usepackage{lipsum}

\usepackage[symbol]{footmisc}
\usepackage{textcomp}

\begin{document}

% \preprint{APS/123-QED}

\title{\texttt{Taweret}: a Python package for Bayesian model mixing}% Force line breaks with \\

\author{K. Ingles}
\email{kingles@illinois.edu}
\affiliation{Illinois Center for Advanced Studies of the Universe \& Department of Physics,\\ University of Illinois Urbana-Champaign, Urbana, IL 61801, USA}

\author{D. Liyanage}
\email{liyanage.1@osu.edu} % change for new place
\affiliation{Department of Physics, The Ohio State University, Columbus, OH 43210, USA} % change for new place

\author{A.~C. Semposki}
\email{as727414@ohio.edu}
\affiliation{Department of Physics and Astronomy and Institute of Nuclear and Particle Physics, Ohio University, Athens, OH 45701, USA}

\author{J.~C. Yannotty}
\email{yannotty.1@buckeyemail.osu.edu}
\affiliation{Department of Statistics, The Ohio State University, Columbus, OH 43210, USA}

\date{\today}% It is always \today, today,
%\maketitle

%\begin{comment}
\begin{center}
  {\LARGE \texttt{Taweret}: a Python package for Bayesian model mixing\par}
  \vspace{2em}
  K. Ingles\textsuperscript{1,*} \footnotetext[1]{kingles@illinois.edu}, D. Liyanage\textsuperscript{2,$\dagger$}\footnotetext[2]{liyanage.1@osu.edu}, A.~C. Semposki\textsuperscript{3,$\ddagger$}\footnotetext[3]{as727414@ohio.edu}, J.~C. Yannotty\textsuperscript{4,$\S$}\footnotetext[4]{yannotty.1@buckeyemail.osu.edu}
  \\
  \vspace{0.5em}
    \textsuperscript{1} 	\textit{\small Illinois Center for Advanced Studies of the Universe \& Department of Physics,\\ University of Illinois Urbana-Champaign, Urbana, IL 61801, USA}  \\
    \textsuperscript{2} \textit{\small Department of Physics, The Ohio State University, Columbus, OH 43210, USA} 
    \\
    \textsuperscript{3} \textit{\small Department of Physics and Astronomy and Institute of Nuclear and Particle Physics, Ohio University, Athens, OH 45701, USA}  
    \\
    \textsuperscript{4} \textit{\small Department of Statistics, The Ohio State University, Columbus, OH 43210, USA}  
    \\
  \vspace{1em}
  \today
\end{center}
%\end{comment}

\section{Summary}
Uncertainty quantification using Bayesian methods is a growing area of research.
Bayesian model mixing (BMM) is a recent development which combines the predictions from multiple models such that each model's best qualities are preserved in the final result. 
Practical tools and analysis suites that facilitate such methods are therefore needed.
\texttt{Taweret} introduces BMM to existing Bayesian uncertainty quantification efforts. Currently \texttt{Taweret} contains three individual Bayesian model mixing techniques, each pertaining to a different type of problem structure; we encourage the future inclusion of user-developed mixing methods.
%Bayesian Model Mixing (BMM) analysis incorporates in parameter estimation, the uncertainty present in selecting one theory or model over another.
\texttt{Taweret}'s first use case is in nuclear physics, but the package has been structured such that it should be adaptable to any research engaged in model comparison or model mixing.
% Nuclear physics is entering the age of uncertainty quantification, both with respect to experimental data and theoretical models.

\section{Statement of need}

In physics applications, multiple models with different physics assumptions can be used to describe an underlying system of interest. It is usually the case that each model has varying fidelity to the observed process across the input domain. Though each model may have similar predictive accuracy on average, the fidelity of the approximation across a subset of the domain may differ drastically for each of the models under consideration. In such cases, inference and prediction based on a single model may be unreliable. One strategy for improving accuracy is to combine, or  ``mix", the predictions from each model using a linear combination or weighted average with input-dependent weights. This approach is intended to improve reliability of inference and prediction and properly quantify model uncertainties. When operating under a Bayesian framework, this technique is referred to as Bayesian model mixing (BMM). In general, model mixing techniques are designed to combine the individual mean predictions or density estimates from the $K$ models under consideration.
For example, \textit{mean-mixing} techniques predict the underlying system by 
\begin{equation}
    E[\bm Y \mid \xvec] = \sum_{k = 1}^K w_k(\xvec)\; f_k(\xvec),
\end{equation}
where $E[\bm Y\mid\xvec]$ denotes the mean of $\bm Y$ given the vector of input parameters $\bm x$, $f_k(\xvec)$ is the mean prediction under the $k^\mathrm{th}$ model $\mathcal{M}_k$, and $w_k(\xvec)$ is the corresponding weight function. The \textit{density-mixing} approach estimates the underlying predictive density by
\begin{equation}
    p(\bm Y_{0} \mid \xvec_{0}) = \sum_{k = 1}^K w_k(\xvec_{0})\;p(\bm Y_{0} \mid \xvec_{0}, \bm{Y}, \M_k),
\end{equation}
where $p(\bm{Y_{0}} \mid \xvec_{0}, \bm{Y}, \M_k)$ represents the predictive density of a future observation $\bm{Y_{0}}$ with respect to the $k^\mathrm{th}$ model $\mathcal{M}_k$ at a new input $\xvec_{0}$. In either BMM setup, a key challenge is defining $w_k(\xvec)$---the functional relationship between the inputs and the weights. 

This work introduces \texttt{Taweret}, a Python package for Bayesian model mixing that includes three novel approaches for combining models, each of which defines the weight function in a unique way (see Table~\ref{tab:methodcomparison} for a comparison of each method). This package has been developed as an integral piece of the Bayesian Analysis of Nuclear Dynamics (BAND) collaboration's software. BAND is a multi-institutional effort to build a cyber-infrastructure framework for use in the nuclear physics community \cite{Phillips:2020dmw, bandframework}. The software is designed to lower the barrier for researchers to employ uncertainty quantification in their experiments, and to integrate, as best as possible, with the community's current standards concerning coding style (\texttt{pep8}). Bayesian model mixing is one of BAND's four central pillars in this framework (the others being emulation, calibration, and experimental design).

In addition to this need, we are aware of several other fields outside of physics that use techniques such as model stacking and Bayesian model averaging (BMA) \cite{Fragoso2018}, e.g., statistics \cite{Yao2018, Yao2022}, meteorology \cite{Sloughter2007}, and neuroscience \cite{FitzGerald2014}. It is expected that the Bayesian model mixing methods presented in \texttt{Taweret} can also be applied to use cases within these fields. 
Statisticians have developed several versatile BMA/stacking packages, e.g. \cite{loo, BMA_R}. However, the only BMM-based package available is \texttt{SAMBA}---a BAND collaboration effort that was developed for testing BMM methods on a toy model \cite{Semposki:2022gcp}. \texttt{Taweret}'s increased functionality, user-friendly structure, and diverse selection of mixing methods make it a marked improvement over \texttt{SAMBA}. 

\section{Structure}

\subsection{Overview of methods}

\begin{table}
    \scriptsize
    \centering
    \begin{tabular}{c|c|c|c|c|c}
    \hline 
    \hline
       Method  & Type & Number of & Number of & Number of & Weight \\
         & & inputs & outputs & models & functions \\
       \hline
       Bivariate linear & Mean \& & &  & & Step,\\
      mixing & Density & 1 & $\geq 1$ & 2 & Sigmoid, \\
       & & & & & Asymmetric 2-step \\
       \hline
       Multivariate mixing & Mean & 1 & 1 & $K$ & Precision \\
       & & & & & weighting\\
       \hline 
       BART mixing & Mean & $\geq 1$ & 1 & $K$ & Regression \\
        & & & & & trees\\
    \hline 
    \hline
    \end{tabular}
    \caption{A summary of the three BMM approaches currently implemented in \texttt{Taweret}. Note that $K\geq 2$. Following the method name and the type of mixing model, the \textit{Number of inputs} column details the dimensions of the parameter which the mixing weights depend on (e.g., in heavy-ion collisions this is the centrality bin); the \textit{Number of outputs} details how many observables the models themselves can have to compute the model likelihood (e.g., in heavy-ion collisions this can include charge multiplicities, transverse momentum distributions, transverse momentum fluctuations, etc.); the \textit{Number of models} column details how many models the mixing method can combine, and the \textit{Weight functions} column describes the available parameterization of how the mixing weights depend on the input parameter.}
    \label{tab:methodcomparison}
\end{table}

\subsubsection{Bivariate linear mixing}

The full description of this mixing method and several of its applications in relativistic heavy-ion collision physics can be found in the Ph.D. thesis of D. Liyanage \cite{Liyanage_thesis}. The bivariate linear mixing method can mix two models either using a density-mixing or a mean-mixing strategy. Currently, this is the only mixing method in \texttt{Taweret} that can also calibrate the models while mixing. It allows the user to choose among the following mixing functions:
\begin{itemize}[nosep]
    \item step: $\Theta(\beta_0-x)$
    \item sigmoid: $\exp\left[(x-\beta_0)/\beta_1\right]$
    \item asymmetric 2-step: $\alpha \Theta(\beta_0-x) + (1-\alpha)\Theta(\beta_1-x)$.
\end{itemize}
Here $\Theta$ denotes the Heaviside step function, $\beta_0$ and $\beta_1$ determine the shape of the weight function and are inferred from the experimental data, and  $x$ is the model input parameter (which is expected to be 1-dimensional for this mixing method).

\subsubsection{Multivariate model mixing} %(A. Semposki's method)

Another Bayesian model mixing method incorporated into \texttt{Taweret} was originally published in \cite{Semposki:2022gcp}, and was the focus of the BMM Python package \texttt{SAMBA} \cite{SAMBA}. It can be described as combining models by weighting each of them by their precision, defined as the inverse of their respective variances. The posterior predictive distribution (PPD) of the mixed model is a Gaussian and can be expressed as
\begin{equation}
    \label{eq:multi_mm_gaussian}
   \mathcal M_\dagger \sim {\mathcal N(f_\dagger, Z_P^{-1})}:
    \quad
    f_{\dagger} = \frac{1}{Z_P}\sum_{k=1}^{K} \frac{1}{\sigma^{2}_k}f_k,
    \quad Z_P \equiv \sum_{k=1}^{K}\frac{1}{\sigma^{2}_k},
\end{equation}
where $\mathcal N(\mu, \sigma^2)$ is a normal distribution with mean $\mu$ and variance $\sigma^2$, $Z_{P}$ is the precision of the models, and each individual model is assumed to possess a Gaussian form such as
\begin{equation}
    {\cal M}_{k} \sim {\cal N}(f_{k}(x),\sigma^2_{k}(x)).
\end{equation}
Here, $f_{k}(x)$ is the mean of the model $k$, and $\sigma^{2}_{k}(x)$ its variance, both at input parameter $x$.
%[[Do we have to say this: The former is usually the stated value of the model at any point in the input space, $x$, and the latter is often an estimation of the model variance. In the case of models that can be described by a series expansion in some parameter, the model error derived from truncated higher order terms can be estimated using the terms one already possesses (see, e.g., \cite{Melendez:2019izc}).]]

In this method, the software receives the one-dimensional input space $X$, the mean of the $K$ models at each point $ x \in X$ (hence it is a mean-based mixing procedure), and the variances of the models at each point in $x \in X$. Each model is assumed to have been calibrated prior to being included in the mix. The ignorance of this mixing method with respect to how each model was generated allows for any model to be used, including Bayesian Machine Learning tools such as Gaussian Processes \cite{Semposki:2022gcp} and Bayesian Neural Networks \cite{Kronheim:2020dmp}.

\subsubsection{Model mixing using Bayesian Additive Regression Trees}
    
A third BMM approach implemented in \texttt{Taweret} adopts a mean-mixing strategy which models the weight functions using Bayesian Additive Regression Trees (BART) conditional on the mean predictions from a set of $K$ models \cite{yannotty2023model}. This approach enables the weight functions to be adaptively learned using tree bases and avoids the need for user-specified basis functions (such as a generalized linear model). Formally, the weight functions are defined by 
\begin{equation}
    w_k(\xvec) = \sum_{j = 1}^m g_k(\xvec; T_j, M_j), \quad \text{for}\ k=1,\ldots,K
\end{equation}
where $g_k(\xvec;T_j,M_j)$ defines the $k^\text{th}$ output of the $j^\text{th}$ tree, $T_j$, using the associated set of parameters, $M_j$. Each weight function is implicitly regularized via a prior to prefer the interval $[0,1]$. Furthermore, the weight functions are not required to sum-to-one and can take values outside of the range of $[0,1]$. This regularization approach is designed to maintain the flexibility of the model while also encouraging the weight functions to take values which preserve desired inferential properties.

This BART-based approach is implemented in \texttt{C++} with the \texttt{trees} module in \texttt{Taweret} acting as a Python interface. The \texttt{C++} back-end implements Bayesian tree models and originates from the \textit{Open Bayesian Trees Project} (\texttt{OpenBT}) \cite{OpenBT_MTP}. This module serves as an example for how existing code bases can be integrated into the \texttt{Taweret} framework.           

\subsection{Overview of package structure}
\begin{figure}
    \centering
    \includegraphics[width=\textwidth]{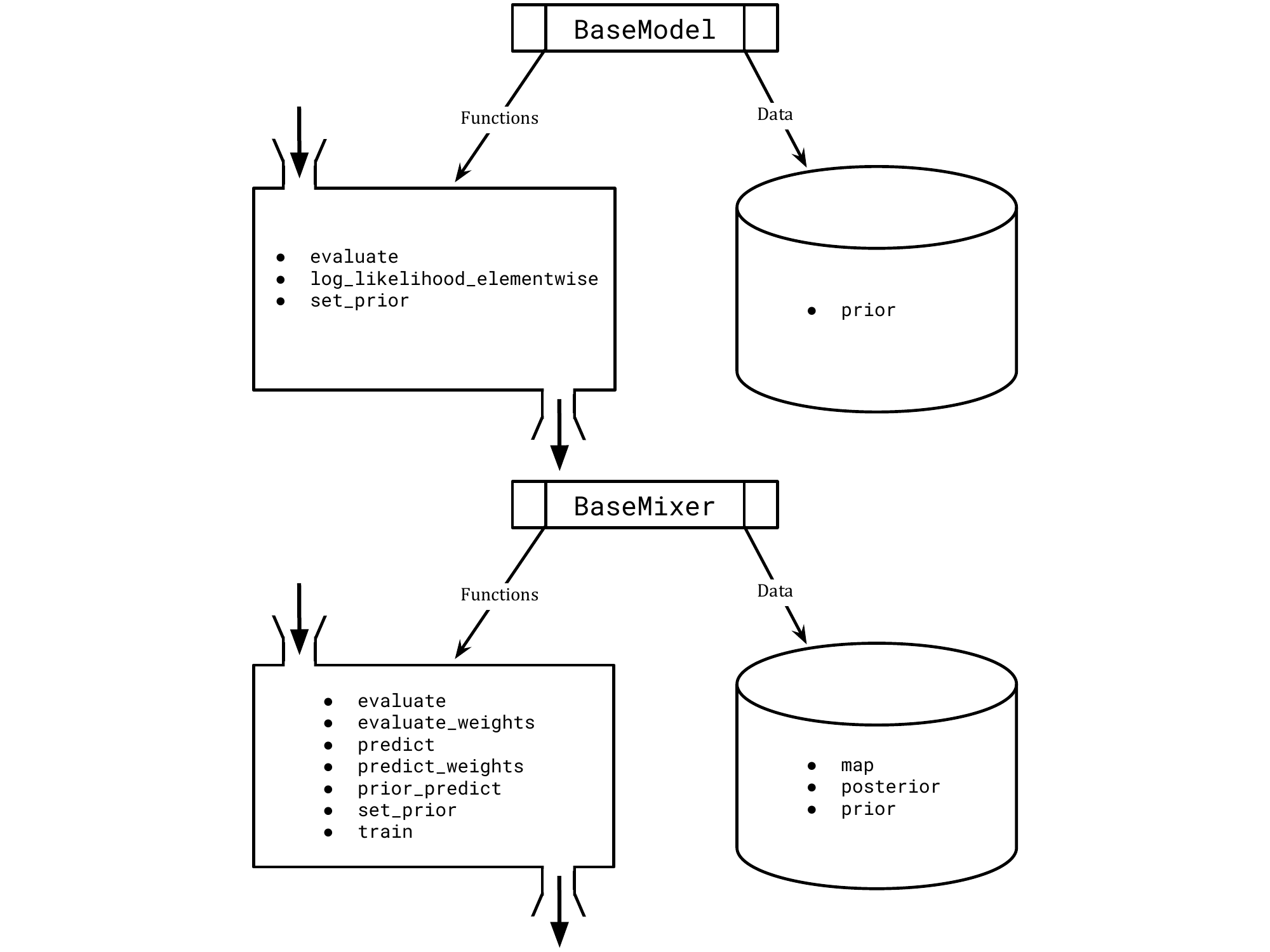}
    \caption{Diagram depicting the base classes, their methods (functions) and their properties (data).}
    \label{fig:codediagram}
\end{figure}

\texttt{Taweret} uses abstract base classes to ensure compatibility and uniformity of models and mixing methods. 
The two base classes are \texttt{BaseModel} and \texttt{BaseMixer} located in the \texttt{core} folder (see Fig.~\ref{fig:codediagram} for a schematic); any model mixing method developed with \texttt{Taweret} is required to inherit from these.
The former represents physics-based models that may include parameters which need to be determined by Bayesian inference.
The latter, \texttt{BaseMixer}, represents a mixing method used to combine the predictions from the physics-based models using Bayesian model mixing.

The design philosophy for \texttt{Taweret} is to make it easy to switch between mixing methods without having to rewrite an analysis script.
Thus, the base classes prescribe which functions need to be present for interoperability between mixing methods, and in particular, the models being called in the method.
The functions required by \texttt{BaseModel} are 
\begin{itemize}[nosep]
    \item \texttt{evaluate} - gives a point prediction for the model;
    \item \texttt{log\_likelihood\_elementwise} - calculates the log-likelihood, reducing along the last axis of an array if the input array has multiple axes;
    \item \texttt{set\_prior} - sets priors for parameters in the model.
\end{itemize}
The functions required by \texttt{BaseMixer} are
\begin{itemize}[nosep]
    \item \texttt{evaluate} - gives point prediction for the mixed model given a set of parameters;
    \item \texttt{evaluate\_weights} - gives point prediction for the weights given a set of parameters;
    \item \texttt{map} - returns the maximum \textit{a posteriori} estimate for the parameters of the mixed model (which includes both the weights and any model parameters);
    \item \texttt{posterior} - returns the chains of the sampled parameters from the mixed model;
    \item \texttt{predict} - returns the posterior predictive distribution for the mixed model;
    \item \texttt{predict\_weights} - returns the posterior predictive distribution for the model weights;
    \item \texttt{prior} - returns the prior distributions (typically objects, not arrays);
    \item \texttt{prior\_predict} - returns the prior predictive distribution for the mixed model;
    \item \texttt{set\_prior} - sets the prior distributions for the mixing method;
    \item \texttt{train} - executes the Bayesian model mixing step.
\end{itemize}

\begin{figure}
    \centering
    \includegraphics[width = 0.95\textwidth, height = 0.45\textwidth]{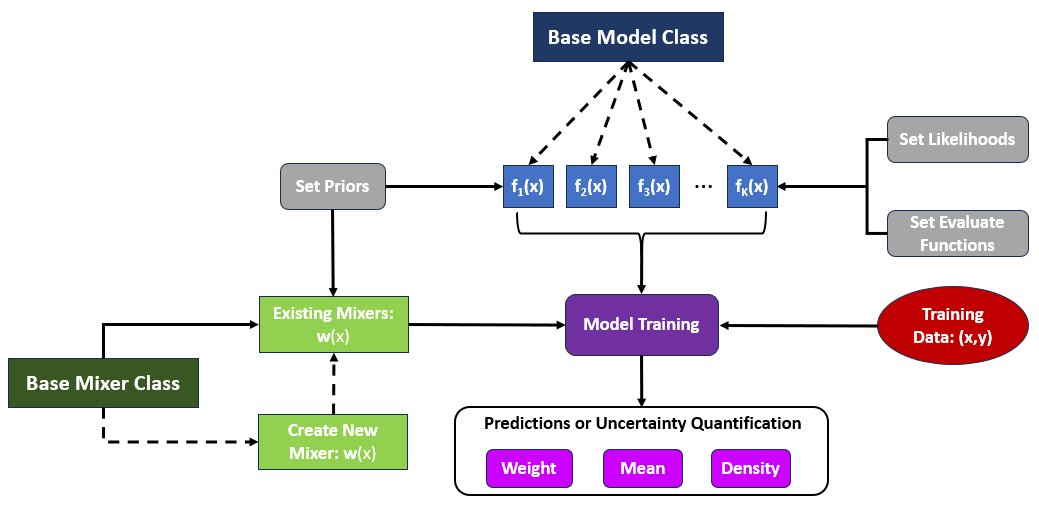}
    \caption{The general workflow for an analysis using \texttt{Taweret}. (Blue) The user must define each of the $K$ models as a class inherited from \texttt{BaseModel}. (Green) The user can select an existing mixing method from \texttt{Taweret} (solid) or contribute a new method (dashed). (Purple) The model is trained using a set of training data (red), the model set (blue), and the selected mixing method (green). Predictions and uncertainty quantification follow from the training process.}
    \label{fig:taweret_workflow}
\end{figure}

Following our design philosophy, the general workflow for an analysis using \texttt{Taweret} is described in Fig.~\ref{fig:taweret_workflow}. From this, one can see three sources of information are generally required for an analysis: a selected mixing method, a model set, and training data. Each of these sources are connected through the training phase, which is where the mixing weights are learned. This leads into the prediction phase, where final predictions are obtained for the overall system and the weight functions. This process is summarized in the code snippet below. This workflow is preserved across the various methods implemented in \texttt{Taweret} and is intended to be maintained for future mixing methods included in this work.

%the setting up and execution of a mixing method should be as simple as 
\begin{verbatim}
```python
    from mix.mix_method import MixMethod
    from models.my_model import MyModel
    
    mixer = MixMethod(models={'model_1': MyModel(...), ...})
    mixer.set_prior(...)
    mixer.train(...)
    mixer.predict(...)
    mixer.predict_weights(...)
'''    
\end{verbatim}
Extending \texttt{Taweret} with a custom class or model simply requires that you inherit from the base classes and implement the required functions. 

\section{Taweret moving forward}

There are certainly many improvements that can be made to \texttt{Taweret}. 
An obvious one is a generalization of the bivariate linear mixing; this could be the mixing of an arbitrary number of models at the density level.
% Such a method would require a stochastic likelihood, i.e. a likelihood that needs to be \emph{sampled} even after the (hyper)parameters have been sampled (e.g., a Dirichlet distribution).
Complementary to this density mixing method is a stochastic, mean-mixing method of arbitrary number $K$ of models.
An extension of the Multivariate Mixing method to multi-dimensional input and output spaces, correlated models, and well as calibration during mixing, is anticipated in future releases.
Lastly, to facilitate the utilization of this growing framework, we hope to enable continuous integration routines for individuals contributing and create docker images that will run \texttt{Taweret}.

\section{Disclosure Statement}

The authors of this work are not aware of any conflicts of interest that would affect this research.

\section{Acknowledgments}

We thank Daniel R. Phillips, Ulrich Heinz, Matt Pratola, Kyle Godbey, Stefan Wild, Sunil Jaiswal, and all other BAND members for crucial feedback and discussion during the development stage of this package. This work is supported by the CSSI program Award OAC-2004601 (DL, ACS, JCY). ACS also acknowledges support from the Department of Energy (contract no. DE-FG02-93ER40756).

\bibliography{references}  % John, Alexandra, BAND framework, etc. included --- need Dan's thesis 

\begin{thebibliography}{15}%
\makeatletter
\providecommand \@ifxundefined [1]{%
 \@ifx{#1\undefined}
}%
\providecommand \@ifnum [1]{%
 \ifnum #1\expandafter \@firstoftwo
 \else \expandafter \@secondoftwo
 \fi
}%
\providecommand \@ifx [1]{%
 \ifx #1\expandafter \@firstoftwo
 \else \expandafter \@secondoftwo
 \fi
}%
\providecommand \natexlab [1]{#1}%
\providecommand \enquote  [1]{``#1''}%
\providecommand \bibnamefont  [1]{#1}%
\providecommand \bibfnamefont [1]{#1}%
\providecommand \citenamefont [1]{#1}%
\providecommand \href@noop [0]{\@secondoftwo}%
\providecommand \href [0]{\begingroup \@sanitize@url \@href}%
\providecommand \@href[1]{\@@startlink{#1}\@@href}%
\providecommand \@@href[1]{\endgroup#1\@@endlink}%
\providecommand \@sanitize@url [0]{\catcode `\\12\catcode `\$12\catcode
  `\&12\catcode `\#12\catcode `\^12\catcode `\_12\catcode `\%12\relax}%
\providecommand \@@startlink[1]{}%
\providecommand \@@endlink[0]{}%
\providecommand \url  [0]{\begingroup\@sanitize@url \@url }%
\providecommand \@url [1]{\endgroup\@href {#1}{\urlprefix }}%
\providecommand \urlprefix  [0]{URL }%
\providecommand \Eprint [0]{\href }%
\providecommand \doibase [0]{https://doi.org/}%
\providecommand \selectlanguage [0]{\@gobble}%
\providecommand \bibinfo  [0]{\@secondoftwo}%
\providecommand \bibfield  [0]{\@secondoftwo}%
\providecommand \translation [1]{[#1]}%
\providecommand \BibitemOpen [0]{}%
\providecommand \bibitemStop [0]{}%
\providecommand \bibitemNoStop [0]{.\EOS\space}%
\providecommand \EOS [0]{\spacefactor3000\relax}%
\providecommand \BibitemShut  [1]{\csname bibitem#1\endcsname}%
\let\auto@bib@innerbib\@empty
%</preamble>
\bibitem [{\citenamefont {Phillips}\ \emph {et~al.}(2021)\citenamefont
  {Phillips}, \citenamefont {Furnstahl}, \citenamefont {Heinz}, \citenamefont
  {Maiti}, \citenamefont {Nazarewicz}, \citenamefont {Nunes}, \citenamefont
  {Plumlee}, \citenamefont {Pratola}, \citenamefont {Pratt}, \citenamefont
  {Viens},\ and\ \citenamefont {Wild}}]{Phillips:2020dmw}%
  \BibitemOpen
  \bibfield  {author} {\bibinfo {author} {\bibfnamefont {D.~R.}\ \bibnamefont
  {Phillips}}, \bibinfo {author} {\bibfnamefont {R.~J.}\ \bibnamefont
  {Furnstahl}}, \bibinfo {author} {\bibfnamefont {U.}~\bibnamefont {Heinz}},
  \bibinfo {author} {\bibfnamefont {T.}~\bibnamefont {Maiti}}, \bibinfo
  {author} {\bibfnamefont {W.}~\bibnamefont {Nazarewicz}}, \bibinfo {author}
  {\bibfnamefont {F.~M.}\ \bibnamefont {Nunes}}, \bibinfo {author}
  {\bibfnamefont {M.}~\bibnamefont {Plumlee}}, \bibinfo {author} {\bibfnamefont
  {M.~T.}\ \bibnamefont {Pratola}}, \bibinfo {author} {\bibfnamefont
  {S.}~\bibnamefont {Pratt}}, \bibinfo {author} {\bibfnamefont {F.~G.}\
  \bibnamefont {Viens}},\ and\ \bibinfo {author} {\bibfnamefont {S.~M.}\
  \bibnamefont {Wild}},\ }\bibfield  {title} {\bibinfo {title} {{Get on the
  BAND Wagon: A Bayesian Framework for Quantifying Model Uncertainties in
  Nuclear Dynamics}},\ }\href {https://doi.org/10.1088/1361-6471/abf1df}
  {\bibfield  {journal} {\bibinfo  {journal} {Journal of Physics G}\ }\textbf
  {\bibinfo {volume} {48}},\ \bibinfo {pages} {072001} (\bibinfo {year}
  {2021})},\ \Eprint {https://arxiv.org/abs/2012.07704} {arXiv:2012.07704
  [nucl-th]} \BibitemShut {NoStop}%
\bibitem [{\citenamefont {Beyer}\ \emph {et~al.}(2023)\citenamefont {Beyer},
  \citenamefont {Buskirk}, \citenamefont {Chan}, \citenamefont {Chang},
  \citenamefont {DeBoer}, \citenamefont {Furnstahl}, \citenamefont {Giuliani},
  \citenamefont {Godbey}, \citenamefont {Ingles}, \citenamefont {Liyanage},
  \citenamefont {Nunes}, \citenamefont {Odell}, \citenamefont {Phillips},
  \citenamefont {Plumlee}, \citenamefont {Pratola}, \citenamefont {Semposki},
  \citenamefont {S\"urer}, \citenamefont {Wild},\ and\ \citenamefont
  {Yannotty}}]{bandframework}%
  \BibitemOpen
  \bibfield  {author} {\bibinfo {author} {\bibfnamefont {K.}~\bibnamefont
  {Beyer}}, \bibinfo {author} {\bibfnamefont {L.}~\bibnamefont {Buskirk}},
  \bibinfo {author} {\bibfnamefont {M.~Y.-H.}\ \bibnamefont {Chan}}, \bibinfo
  {author} {\bibfnamefont {T.~H.}\ \bibnamefont {Chang}}, \bibinfo {author}
  {\bibfnamefont {R.~J.}\ \bibnamefont {DeBoer}}, \bibinfo {author}
  {\bibfnamefont {R.~J.}\ \bibnamefont {Furnstahl}}, \bibinfo {author}
  {\bibfnamefont {P.}~\bibnamefont {Giuliani}}, \bibinfo {author}
  {\bibfnamefont {K.}~\bibnamefont {Godbey}}, \bibinfo {author} {\bibfnamefont
  {K.}~\bibnamefont {Ingles}}, \bibinfo {author} {\bibfnamefont
  {D.}~\bibnamefont {Liyanage}}, \bibinfo {author} {\bibfnamefont {F.~M.}\
  \bibnamefont {Nunes}}, \bibinfo {author} {\bibfnamefont {D.}~\bibnamefont
  {Odell}}, \bibinfo {author} {\bibfnamefont {D.~R.}\ \bibnamefont {Phillips}},
  \bibinfo {author} {\bibfnamefont {M.}~\bibnamefont {Plumlee}}, \bibinfo
  {author} {\bibfnamefont {M.~T.}\ \bibnamefont {Pratola}}, \bibinfo {author}
  {\bibfnamefont {A.~C.}\ \bibnamefont {Semposki}}, \bibinfo {author}
  {\bibfnamefont {O.}~\bibnamefont {S\"urer}}, \bibinfo {author} {\bibfnamefont
  {S.~M.}\ \bibnamefont {Wild}},\ and\ \bibinfo {author} {\bibfnamefont
  {J.~C.}\ \bibnamefont {Yannotty}},\ }\href
  {https://github.com/bandframework/bandframework} {\emph {\bibinfo {title}
  {{BANDFramework: An} Open-Source Framework for {B}ayesian Analysis of Nuclear
  Dynamics}}},\ \bibinfo {type} {Tech. Rep.}\ \bibinfo {number} {Version
  0.3.0}\ (\bibinfo {year} {2023})\BibitemShut {NoStop}%
\bibitem [{\citenamefont {Fragoso}\ \emph {et~al.}(2018)\citenamefont
  {Fragoso}, \citenamefont {Bertoli},\ and\ \citenamefont
  {Louzada}}]{Fragoso2018}%
  \BibitemOpen
  \bibfield  {author} {\bibinfo {author} {\bibfnamefont {T.~M.}\ \bibnamefont
  {Fragoso}}, \bibinfo {author} {\bibfnamefont {W.}~\bibnamefont {Bertoli}},\
  and\ \bibinfo {author} {\bibfnamefont {F.}~\bibnamefont {Louzada}},\
  }\bibfield  {title} {\bibinfo {title} {Bayesian model averaging: A systematic
  review and conceptual classification},\ }\href
  {https://doi.org/https://doi.org/10.1111/insr.12243} {\bibfield  {journal}
  {\bibinfo  {journal} {International Statistical Review}\ }\textbf {\bibinfo
  {volume} {86}},\ \bibinfo {pages} {1} (\bibinfo {year} {2018})},\ \Eprint
  {https://arxiv.org/abs/https://onlinelibrary.wiley.com/doi/pdf/10.1111/insr.12243}
  {https://onlinelibrary.wiley.com/doi/pdf/10.1111/insr.12243} \BibitemShut
  {NoStop}%
\bibitem [{\citenamefont {Yao}\ \emph {et~al.}(2018)\citenamefont {Yao},
  \citenamefont {Vehtari}, \citenamefont {Simpson},\ and\ \citenamefont
  {Gelman}}]{Yao2018}%
  \BibitemOpen
  \bibfield  {author} {\bibinfo {author} {\bibfnamefont {Y.}~\bibnamefont
  {Yao}}, \bibinfo {author} {\bibfnamefont {A.}~\bibnamefont {Vehtari}},
  \bibinfo {author} {\bibfnamefont {D.}~\bibnamefont {Simpson}},\ and\ \bibinfo
  {author} {\bibfnamefont {A.}~\bibnamefont {Gelman}},\ }\bibfield  {title}
  {\bibinfo {title} {Using stacking to average bayesian predictive
  distributions (with discussion)},\ }\bibfield  {journal} {\bibinfo  {journal}
  {Bayesian Analysis, 13 (3) 917 - 1007}\ }\href
  {https://doi.org/10.1214/17-BA1091} {10.1214/17-BA1091} (\bibinfo {year}
  {2018})\BibitemShut {NoStop}%
\bibitem [{\citenamefont {Yao}\ \emph {et~al.}(2022)\citenamefont {Yao},
  \citenamefont {Pirš}, \citenamefont {Vehtari},\ and\ \citenamefont
  {Gelman}}]{Yao2022}%
  \BibitemOpen
  \bibfield  {author} {\bibinfo {author} {\bibfnamefont {Y.}~\bibnamefont
  {Yao}}, \bibinfo {author} {\bibfnamefont {G.}~\bibnamefont {Pirš}}, \bibinfo
  {author} {\bibfnamefont {A.}~\bibnamefont {Vehtari}},\ and\ \bibinfo {author}
  {\bibfnamefont {A.}~\bibnamefont {Gelman}},\ }\bibfield  {title} {\bibinfo
  {title} {{Bayesian Hierarchical Stacking: Some Models Are (Somewhere)
  Useful}},\ }\href {https://doi.org/10.1214/21-BA1287} {\bibfield  {journal}
  {\bibinfo  {journal} {Bayesian Analysis}\ }\textbf {\bibinfo {volume} {17}},\
  \bibinfo {pages} {1043 } (\bibinfo {year} {2022})}\BibitemShut {NoStop}%
\bibitem [{\citenamefont {Sloughter}\ \emph {et~al.}(2007)\citenamefont
  {Sloughter}, \citenamefont {Raftery}, \citenamefont {Gneiting},\ and\
  \citenamefont {Fraley}}]{Sloughter2007}%
  \BibitemOpen
  \bibfield  {author} {\bibinfo {author} {\bibfnamefont {J.~M.}\ \bibnamefont
  {Sloughter}}, \bibinfo {author} {\bibfnamefont {A.~E.}\ \bibnamefont
  {Raftery}}, \bibinfo {author} {\bibfnamefont {T.}~\bibnamefont {Gneiting}},\
  and\ \bibinfo {author} {\bibfnamefont {C.}~\bibnamefont {Fraley}},\
  }\bibfield  {title} {\bibinfo {title} {Probabilistic quantitative
  precipitation forecasting using {B}ayesian model averaging},\ }\bibfield
  {journal} {\bibinfo  {journal} {Monthly Weather Review, 135, 3209–3220}\
  }\href {https://doi.org/10.1175/MWR3441.1} {10.1175/MWR3441.1} (\bibinfo
  {year} {2007})\BibitemShut {NoStop}%
\bibitem [{\citenamefont {FitzGerald}\ \emph {et~al.}(2014)\citenamefont
  {FitzGerald}, \citenamefont {Dolan},\ and\ \citenamefont
  {Friston}}]{FitzGerald2014}%
  \BibitemOpen
  \bibfield  {author} {\bibinfo {author} {\bibfnamefont {T.~H.~B.}\
  \bibnamefont {FitzGerald}}, \bibinfo {author} {\bibfnamefont {R.~J.}\
  \bibnamefont {Dolan}},\ and\ \bibinfo {author} {\bibfnamefont {K.~J.}\
  \bibnamefont {Friston}},\ }\bibfield  {title} {\bibinfo {title} {Model
  averaging, optimal inference, and habit formation},\ }\bibfield  {journal}
  {\bibinfo  {journal} {Frontiers in Human Neuroscience, 8:457}\ }\href
  {https://doi.org/10.3389/fnhum.2014.00457} {10.3389/fnhum.2014.00457}
  (\bibinfo {year} {2014})\BibitemShut {NoStop}%
\bibitem [{\citenamefont {Vehtari}\ \emph {et~al.}(2017)\citenamefont
  {Vehtari}, \citenamefont {Gelman},\ and\ \citenamefont {Gabry}}]{loo}%
  \BibitemOpen
  \bibfield  {author} {\bibinfo {author} {\bibfnamefont {A.}~\bibnamefont
  {Vehtari}}, \bibinfo {author} {\bibfnamefont {A.}~\bibnamefont {Gelman}},\
  and\ \bibinfo {author} {\bibfnamefont {J.}~\bibnamefont {Gabry}},\ }\bibfield
   {title} {\bibinfo {title} {\texttt{loo}: Efficient leave-one-out
  cross-validation and waic for {B}ayesian models}\ }(\bibinfo {year} {2017})\
  \bibinfo {note} {\mbox{https://mc-stan.org/loo/}}\BibitemShut {NoStop}%
\bibitem [{\citenamefont {Raftery}\ \emph {et~al.}(2022)\citenamefont
  {Raftery}, \citenamefont {Hoeting}, \citenamefont {Volinsky}, \citenamefont
  {Painter},\ and\ \citenamefont {Yeung}}]{BMA_R}%
  \BibitemOpen
  \bibfield  {author} {\bibinfo {author} {\bibfnamefont {A.}~\bibnamefont
  {Raftery}}, \bibinfo {author} {\bibfnamefont {J.}~\bibnamefont {Hoeting}},
  \bibinfo {author} {\bibfnamefont {C.}~\bibnamefont {Volinsky}}, \bibinfo
  {author} {\bibfnamefont {I.}~\bibnamefont {Painter}},\ and\ \bibinfo {author}
  {\bibfnamefont {K.~Y.}\ \bibnamefont {Yeung}},\ }\href
  {https://CRAN.R-project.org/package=BMA} {\emph {\bibinfo {title} {BMA:
  Bayesian Model Averaging}}} (\bibinfo {year} {2022}),\ \bibinfo {note} {r
  package version 3.18.17}\BibitemShut {NoStop}%
\bibitem [{\citenamefont {Semposki}\ \emph
  {et~al.}(2022{\natexlab{a}})\citenamefont {Semposki}, \citenamefont
  {Furnstahl},\ and\ \citenamefont {Phillips}}]{Semposki:2022gcp}%
  \BibitemOpen
  \bibfield  {author} {\bibinfo {author} {\bibfnamefont {A.~C.}\ \bibnamefont
  {Semposki}}, \bibinfo {author} {\bibfnamefont {R.~J.}\ \bibnamefont
  {Furnstahl}},\ and\ \bibinfo {author} {\bibfnamefont {D.~R.}\ \bibnamefont
  {Phillips}},\ }\bibfield  {title} {\bibinfo {title} {{Interpolating between
  small- and large-g expansions using Bayesian model mixing}},\ }\href
  {https://doi.org/10.1103/PhysRevC.106.044002} {\bibfield  {journal} {\bibinfo
   {journal} {Physical Review C}\ }\textbf {\bibinfo {volume} {106}},\ \bibinfo
  {pages} {044002} (\bibinfo {year} {2022}{\natexlab{a}})},\ \Eprint
  {https://arxiv.org/abs/2206.04116} {arXiv:2206.04116} \BibitemShut {NoStop}%
\bibitem [{\citenamefont {Liyanage}(2023)}]{Liyanage_thesis}%
  \BibitemOpen
  \bibfield  {author} {\bibinfo {author} {\bibfnamefont {D.}~\bibnamefont
  {Liyanage}},\ }\emph {\bibinfo {title} {Multifaceted study of
  ultrarelativistic heavy ion collisions}},\ \href
  {http://rave.ohiolink.edu/etdc/view?acc_num=osu1689508609203123} {Ph.D.
  thesis},\ \bibinfo  {school} {The Ohio State University} (\bibinfo {year}
  {2023})\BibitemShut {NoStop}%
\bibitem [{\citenamefont {Semposki}\ \emph
  {et~al.}(2022{\natexlab{b}})\citenamefont {Semposki}, \citenamefont
  {Furnstahl},\ and\ \citenamefont {Phillips}}]{SAMBA}%
  \BibitemOpen
  \bibfield  {author} {\bibinfo {author} {\bibfnamefont {A.~C.}\ \bibnamefont
  {Semposki}}, \bibinfo {author} {\bibfnamefont {R.~J.}\ \bibnamefont
  {Furnstahl}},\ and\ \bibinfo {author} {\bibfnamefont {D.~R.}\ \bibnamefont
  {Phillips}},\ }\bibfield  {title} {\bibinfo {title} {{{SAMBA: SAndbox for
  Mixing using Bayesian Analysis}}}\ }(\bibinfo {year} {2022})\ \bibinfo {note}
  {\url{https://github.com/asemposki/SAMBA}}\BibitemShut {NoStop}%
\bibitem [{\citenamefont {Kronheim}\ \emph {et~al.}(2022)\citenamefont
  {Kronheim}, \citenamefont {Kuchera},\ and\ \citenamefont
  {Prosper}}]{Kronheim:2020dmp}%
  \BibitemOpen
  \bibfield  {author} {\bibinfo {author} {\bibfnamefont {B.}~\bibnamefont
  {Kronheim}}, \bibinfo {author} {\bibfnamefont {M.}~\bibnamefont {Kuchera}},\
  and\ \bibinfo {author} {\bibfnamefont {H.}~\bibnamefont {Prosper}},\
  }\bibfield  {title} {\bibinfo {title} {{TensorBNN: Bayesian inference for
  neural networks using TensorFlow}},\ }\href
  {https://doi.org/10.1016/j.cpc.2021.108168} {\bibfield  {journal} {\bibinfo
  {journal} {Computer Physics Communications}\ }\textbf {\bibinfo {volume}
  {270}},\ \bibinfo {pages} {108168} (\bibinfo {year} {2022})},\ \Eprint
  {https://arxiv.org/abs/2009.14393} {arXiv:2009.14393 [physics.comp-ph]}
  \BibitemShut {NoStop}%
\bibitem [{\citenamefont {Yannotty}\ \emph {et~al.}(2023)\citenamefont
  {Yannotty}, \citenamefont {Santner}, \citenamefont {Furnstahl},\ and\
  \citenamefont {Pratola}}]{yannotty2023model}%
  \BibitemOpen
  \bibfield  {author} {\bibinfo {author} {\bibfnamefont {J.~C.}\ \bibnamefont
  {Yannotty}}, \bibinfo {author} {\bibfnamefont {T.~J.}\ \bibnamefont
  {Santner}}, \bibinfo {author} {\bibfnamefont {R.~J.}\ \bibnamefont
  {Furnstahl}},\ and\ \bibinfo {author} {\bibfnamefont {M.~T.}\ \bibnamefont
  {Pratola}},\ }\href {https://doi.org/10.1080/00401706.2023.2257765} {\bibinfo
  {title} {Model mixing using {B}ayesian additive regression trees}} (\bibinfo
  {year} {2023})\BibitemShut {NoStop}%
\bibitem [{\citenamefont {Pratola}\ \emph {et~al.}(2023)\citenamefont
  {Pratola}, \citenamefont {McCulloch}, \citenamefont {Chipman},\ and\
  \citenamefont {Horiguchi}}]{OpenBT_MTP}%
  \BibitemOpen
  \bibfield  {author} {\bibinfo {author} {\bibfnamefont {M.~T.}\ \bibnamefont
  {Pratola}}, \bibinfo {author} {\bibfnamefont {R.~E.}\ \bibnamefont
  {McCulloch}}, \bibinfo {author} {\bibfnamefont {H.~A.}\ \bibnamefont
  {Chipman}},\ and\ \bibinfo {author} {\bibfnamefont {A.}~\bibnamefont
  {Horiguchi}},\ }\href {https://bitbucket.org/mpratola/openbt/wiki/Home}
  {\emph {\bibinfo {title} {{Open Bayesian Trees Project}}}} (\bibinfo {year}
  {2023})\BibitemShut {NoStop}%
\end{thebibliography}%

\end{document}